\documentclass[letterpaper, 10 pt, conference]{ieeeconf}  
\usepackage[backend=biber,style=numeric,sorting=none,maxnames=2,minnames=1]{biblatex}
\usepackage{graphicx}
\usepackage{color, soul}
\usepackage{subcaption}
\usepackage{amsmath}
\usepackage{multirow}
\usepackage{booktabs}
\usepackage{algorithm}
\usepackage{algorithmicx}
\usepackage{algpseudocode} 
\usepackage{float}
\usepackage{tikz}

\newcommand\copyrighttext{%
  \footnotesize \textcopyright~2025 IEEE. Personal use of this material is permitted. Permission from IEEE must be obtained for all other uses, in any current or future media,
including reprinting/republishing this material for advertising or promotional purposes, creating new collective works, for resale or redistribution to servers
or lists, or reuse of any copyrighted component of this work in other works.}

\newcommand\copyrightnotice{%
    \begin{tikzpicture}[remember picture,overlay]
    \node[anchor=south,yshift=10pt] at (current page.south) {\fbox{\parbox{\dimexpr\textwidth-\fboxsep-\fboxrule\relax}{\copyrighttext}}};
    \end{tikzpicture}%
}

\bibliography{literature}
\AtEveryBibitem{\small
  \setlength{\itemsep}{0pt}
  \setlength{\parskip}{0pt}
  \setlength{\bibitemsep}{0pt plus 0.3ex}
  \setlength{\bibitemsep}{0pt}
  \setlength{\bibhang}{0pt}
  \setlength{\bibnamesep}{0pt}
  \setlength{\biblabelsep}{3pt}
}

\IEEEoverridecommandlockouts                              
\title{\LARGE \bf
Generating Automotive Code: Large Language Models for Software Development and Verification in Safety-Critical Systems

\author{Sven Kirchner$^{1}$, Alois C. Knoll$^{1}$}
\thanks{\hspace*{-1em}$^{1}$ Technical University of Munich, Garching, Bayern, Germany}%
\thanks{\hrule}
\thanks{\hspace*{-1em}This research was funded by the Federal Ministry of Education and Research of Germany (BMBF) as part of the CeCaS project, FKZ: 16ME0800K.}
}

\begin{document}

\makeatother
\maketitle
\copyrightnotice
\thispagestyle{empty}
\pagestyle{empty}


\begin{abstract}
Developing safety-critical automotive software presents significant challenges due to increasing system complexity and strict regulatory demands. This paper proposes a novel framework integrating Generative Artificial Intelligence (GenAI) into the Software Development Lifecycle (SDLC). The framework uses Large Language Models (LLMs) to automate code generation in languages such as C++, incorporating safety-focused practices such as static verification, test-driven development and iterative refinement. A feedback-driven pipeline ensures the integration of test, simulation and verification for compliance with safety standards. The framework is validated through the development of an Adaptive Cruise Control (ACC) system. Comparative benchmarking of LLMs ensures optimal model selection for accuracy and reliability. Results demonstrate that the framework enables automatic code generation while ensuring compliance with safety-critical requirements, systematically integrating GenAI into automotive software engineering. This work advances the use of AI in safety-critical domains, bridging the gap between state-of-the-art generative models and real-world safety requirements.
\end{abstract}


\section{INTRODUCTION}
Recent advancements in the automotive domain are driving a paradigm shift from hardware-defined to software-defined intelligent vehicles, where software complexity and safety-criticality have increased significantly. Traditional linear processes, such as the V-model or the waterfall model \cite{pressman2009}, offer limited flexibility to adapt to dynamically evolving requirements. As the volume of automotive software grows, each change in requirements requires extensive changes to the code base and repeated validation cycles, increasing both development time and cost. As a result, software architects and developers face increasing challenges in ensuring safety compliance, especially given the continuous expansion of regulatory frameworks and standards, which are now reaching levels of complexity that are difficult to track and implement manually \cite{serban2018}. Large Language Models (LLMs), such as Chat GPT-3 \cite{gpt2020}, have recently shown promise in addressing this complexity by transforming the role of developers from code authors to orchestrators of generative pipelines. Instead of writing all application-level software manually, engineers can leverage LLMs for automated code generation, using the same standards that once hindered rapid development as structured data sources for compliance. 

In this work, we propose a novel framework that integrates Generative Artificial Intelligence (GenAI) into the software development lifecycle (SDLC). By using LLMs in conjunction with test-driven development (TDD) and static analysis, our approach enables modular system architectures that can be rapidly adapted to evolving requirements, while ensuring compliance with critical safety standards. A core element of the proposed framework is its focus on software generation during the test and integration phases, making the LLM an active participant in the iterative refinement loop. Under this paradigm, test suites and integration scripts assist the LLM by guiding automated code generation to meet specified requirements. The automated process reduces the need for manual recoding and retesting when system requirements change. Our methodology therefore shifts engineering effort to the creation of specification artefacts and robust tools, rather than traditional manual coding. We detail how this framework improves development speed by minimizing human intervention in code production and compliance checks and we illustrate its capabilities with an automotive case study that demonstrates its ability to save time and reduce error rates. We therefore present the following \textbf{contributions}:

\begin{itemize} 
    \item \textbf{GenAI-Integrated SDLC:} A novel LLM-driven development cycle that combines TDD, static analysis and iterative refinement for safety-critical automotive software. 
    \item \textbf{Safety Monitoring Pipeline:} A unified framework for static analysis, formal verification, and automated integration validation to ensure safety compliance in automotive software systems. 
    \item \textbf{LLM Handling: }Evaluating the benchmark performance of LLMs and optimizing their implementation for automated code generation and refinement in automotive software development.
    \item \textbf{ISO-based ACC Case Study:} Validation through automated generation of an ISO-based ACC system in C++, tested on the CARLA simulator \cite{dosovitskiy2017}. 
\end{itemize}


\section{Related Work}

\begin{figure*}[htb]
    \vspace{1em}
    \centering
    \includegraphics[width=0.96\textwidth]{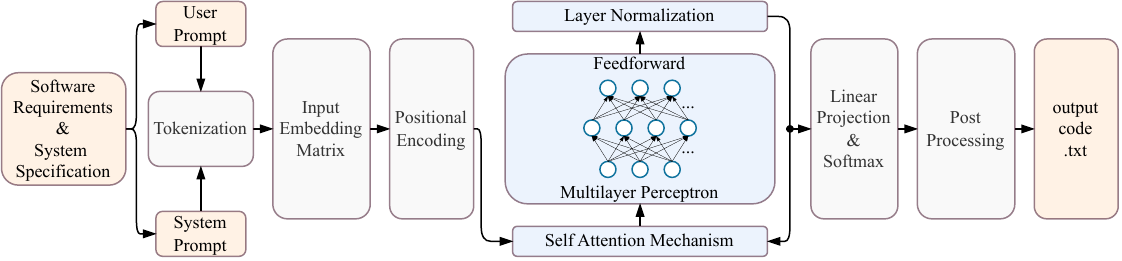}
    \caption{Introduction of Attention Mechanism and Transformer Architecture in Large Language Models for Code Generation.}
    \label{fig:Transformer_Architecture}
\end{figure*}

The fundamental principles of software engineering are based on critical design decisions in software development. To integrate safety as a priority, a focus on robust design capabilities can be complemented by effective test and validation methodologies and the use of software verification tools. This enables GenAI to create scalable and safety-critical applications while ensuring compliance with automotive standards.

\subsection{Reducing software complexity through design choices}
Managing complexity in software engineering requires a systematic approach to design choices, aimed at constraining degrees of freedom and thereby reducing the likelihood of errors. Consequently, maintaining logical consistency in algorithms and flow control becomes essential for verifying correctness \cite{sommerville2011}. Structured interactions between the system and its environment reduce ambiguity and improve integration \cite{pressman2014}. Efficient organization of data underpins performance and scalability \cite{silberschatz2020}, while modular architectures ensure extensibility and adaptability \cite{bass2012}. Test-driven development integrates correctness into the development process \cite{beck2003}, while computational precision mitigates numerical instability \cite{higham2002}. Dependency management and platform compatibility ensure consistent behaviour across environments \cite{fowler2004}. Addressing security vulnerabilities is essential for demonstrating system safety \cite{viega2001}, while effectively managing concurrency is critical for handling dynamic and parallel systems \cite{shavit2012}.

Design principles and standards provide a formal framework for managing the inherent degrees of freedom in software development, ultimately enabling the creation of robust, scalable and extensible systems \cite{bass2012}.
Strategically controlling each degree of freedom minimizes the potential for error, thereby increasing the overall safety and reliability of the software.

\subsection{Safety-Critical Software} 
Safety-critical software requires a systematic approach, treating programming as an exact science with predictable and provable behaviour under all conditions \cite{Dijkstra1997}. Achieving this requires careful selection of programming languages, robust compiler validation and comprehensive verification and testing methodologies.

C++ is widely used in safety-critical design because of to its balance of high performance, precise memory control, and deterministic resource management, which enables strict real-time and reliability constraints to be met. Compiler validation further strengthens these guarantees by translating the code correctly. C++ compilers such as GCC and Clang are highly optimized for performance and reliability, offering advanced static analysis, code optimization and diagnostics \cite{Stroustrup2013}. Beyond language and compiler choice, static code analysis is essential to ensure logical consistency \cite{sommerville2011}. Tools such as cppcheck for C++ \cite{cppcheck2024} identify memory leaks, race conditions and guideline violations, significantly reducing the likelihood of unexpected behaviors. Before full integration, system behaviour is validated through unit testing. Frameworks like Google Test \cite{GoogleTest} verify functional correctness by adapting testing preconditions and edge cases. By integrating language safety, certified compilers, static analysis and rigorous testing, this approach improves correctness, compliance with safety standards and robustness in safety-critical applications.

\begin{figure*}[htb]
    \vspace{0.5cm}
    \centering
    \includegraphics[width=0.96\textwidth]{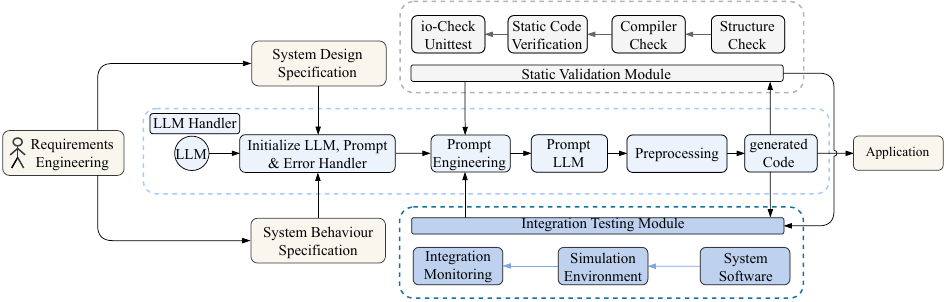}
    \caption{The code generation architecture consists of three components: the LLM handler (light blue), the static validation module (grey) and the integration testing module (dark blue). The user (light yellow) provides specifications and the LLM Handler generates executable code that is iteratively refined by static checks and integration tests to ensure safety and functionality.}
    \label{fig:architecture}
\end{figure*}

\subsection{Foundations of Requirements Engineering and Software Development for Automotive Systems}  
Automotive software development bridges complex safety and functional requirements with robust code design. A Software Requirements Specification (SRS) defines both functional (e.g., system behaviour) and non-functional (e.g., performance, security) requirements, guiding the development process. Stakeholder alignment is achieved through use cases, user stories and prototypes, ensuring clarity and addressing safety from the outset.
Safety and quality standards are central. ISO 26262 \cite{ISO26262} defines functional safety requirements for road vehicles, while Automotive Software Process Improvement and Capability Determination (ASPICE) \cite{ASPICE2022} provides a framework for assessing software quality and processes. The MISRA guidelines \cite{MISRA2012} standardize programming practices, often implemented in C or C++ for their reliability and performance in safety-critical systems. Detailed testing, including static analysis and unit testing, ensures compliance with safety standards. General requirements align with frameworks like ISO 26262, while function-specific requirements address unique system needs.

By integrating these principles, automotive software development transforms complex requirements into reliable, maintainable and safety-compliant systems. 

\subsection{Integrating Generative AI into Software Engineering}
The introduction of the Attention Mechanism revolutionized natural language processing, enabling the development of Large Language Models (LLMs) through the innovative Transformer Architecture. This architecture facilitates non-sequential data processing, overcoming the limitations of earlier recurrent models and introducing greater efficiency and scalability \cite{vaswani2023}.

In the framework illustrated in Figure \ref{fig:Transformer_Architecture}, a given requirement provided as textual input is tokenized and processed through input embeddings, where it is enriched with positional encodings to preserve the sequential context. The tokens are then iteratively passed through a stack of self-attention layers and feedforward multilayer perceptrons. These mechanisms ensure that the model captures both local and global dependencies within the input requirements. Subsequently, the processed tokens undergo linear projection and post-processing steps to generate the next token in the sequence. This iterative process enables the generation of software in the form of coherent and contextually relevant text. The ability to train or fine-tune foundation models like llama3 \cite{grattafiori2024} on specific tasks has been further enhanced by leveraging large-scale datasets. For instance, LLMs such as those designed for code generation like Qwen2.5-Coder \cite{hui2024} benefit significantly from task-specific training datasets. The performance of LLMs is heavily influenced by the quality of the training data, the design of the prompt and the system specifications.

An essential aspect of the effective use of LLMs is the management of the context size, which is inherently limited by the architecture. To maximize the utility of the available context, effective prompting techniques have been developed, ensuring that critical information is succinctly presented within the limited input space. Strategies such as Zero-shot or Few-shot Prompting \cite{Scius2024}, Chain-of-Thought \cite{zhang2024} and Role-based Prompting \cite{kong2024} enable LLMS to generate accurate and high quality output for a variety of tasks. Combination with further validation and formal verification methods can improve the overall quality of the generated code \cite{sevenhuijsen2025}.


\section{Method}
To integrate generative AI into the software development cycle, we propose an approach that combines test-driven development with previously introduced verification methods and static code analysis, ensuring safety through rapid feedback and consistency.

\subsection{Architectural Design and Software Version Control}
The framework illustrated in Figure \ref{fig:architecture}) shows the GenAI integrated SDLC. User input, including detailed specifications, serves as the foundation. The LLM Handler transforms this input into executable code and iteratively refines it based on feedback from subsequent phases. The Static Validation Module analyzes the generated code for compliance with safety standards and design principles. Finally, the Integration Testing Module evaluates the system in a dynamic test environment, ensuring robust performance and functional correctness.

\begin{figure}[h]
    \vspace{1em}
    \centering
    \includegraphics[width=0.4\textwidth]{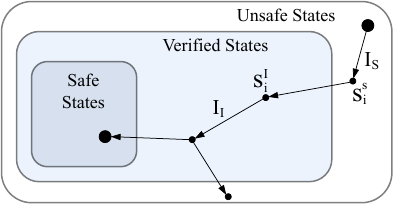}
    \caption{Software versions are categorized by safety classification, progressing through three primary states. The "Verified State" is achieved upon successful completion of static testing. "Safe States" are achieved after meeting static and integration test criteria. The LLM Handler dynamically manages the generation of LLMs based on the current state of the software.}
    \label{fig:safe_states}
\end{figure}

The use of automated code refinement and regeneration is highly dependent on the software's current development stage and version. In safety-critical systems, version control adheres to a methodology combining integration monitoring, static code analysis and iterative refinement (Figure \ref{fig:safe_states}). The ultimate objective is to achieve a "Safe State," wherein all predefined functional and safety requirements are satisfied.

Each iteration \(I_S\) involves static analysis managed by the Static Validation module to detect and resolve logical inconsistencies and design violations. Upon successful completion, the code moves from a static state ($s^S_i$) to an integration state ($s^I_i$) within a simulation environment where iterative validation and refinement occurs (\(I_I\)).  Transition to a safe state is only realized when both static and integration criteria are conclusively met.

This cyclical process of static and integration iterations ensures continuous improvement of the software. Each subsequent state builds incrementally upon its predecessor, anchored in validated safety and integration protocols. Importantly, a new verified state only generated when the integration phase is successful and all static checks are resolved, ensuring an uncompromising commitment to safety and correctness.

\subsection{LLM based generation: Specification and User Input}
User input is given in two distinct classes: System Design Specification and System Behaviour Specification. The System Design Specification focuses on the inputs and outputs of the system, using precise mathematical language to define algorithmic preconditions and postconditions. The System Behaviour Specification describes the overall structure and behaviour of the system.
For prompting LLMs, structured text is provided as input. JSON is a widely used in software engineering due to its standardization and compatibility. YAML's human-readable features, such as whitespace structuring, optional quotes and support for inline comments, enhance usability and interpretability during the specification phase \cite{Eriksson2011}. To maximize prompt efficiency,  JSON is used for the system design specification, while YAML is used for the system behaviour specification. This dual-format strategy ensures effective context management and optimal use of prompt space to generate accurate and interpretable output.

The framework integrates zero-shot and few-shot prompting to optimize LLM performance. Zero-shot prompting establishes baseline output, followed by iterative few-shot refinement to improve accuracy and resolve errors. This process transitions from exploratory prompts to precise adjustments based on output quality. Chain-of-Thought reasoning improves version control by providing the LLM with the best prior solution, enabling iterative improvement towards a safe state, as shown in Figure~\ref{fig:safe_states}. Role-based prompting defines the LLM’s role as a "specialized AI assistant for safety-critical automotive code generation", ensuring outputs are tailored to the framework's requirements. Each iteration builds on previous results, driving continuous improvement and alignment with specifications.

Selecting an appropriate LLM is critical to achieving optimal results. We use McEval: Massively Multilingual Code Evaluation \cite{mceval2024}, a benchmark designed to evaluate LLM performance across 40 programming languages, including Rust and C++, using 16,000 test cases. This provides a comprehensive framework for evaluating multilingual code generation. For reasoning and iterative code refinement, we adopt Aider's code editing benchmark \cite{aider2024}. It evaluates precision and consistency in modifying functions, implementing missing functionality and refactoring code from natural language instructions. By prioritizing editing over generation , the Aider benchmark evaluates the accuracy and consistency of code review and refinement in a variety of programming challenges, making it critical for frameworks that require robust iterative development capabilities. We also include the widely recognized HumanEval benchmark \cite{chen2021}. Table \ref{tab:LLM_benchmarking} summarizes the performance of various large language models across these benchmarks, highlighting their multilingual code generation and iterative refinement capabilities. We evaluate state-of-the-art open-source LLMs, including Qwen2.5-Coder-7B-Instruct, Llama-3-8B-Instruct, DeepSeek-Coder-V2 Lite Instruct \cite{deepseekai2024}, DeepSeek-Coder 33B Instruct \cite{guo2024}, and CodeStral-22B \cite{codestral2024}, using GPT-4o as a benchmark for comparison.

\begin{table}[h]
\centering
\begin{tabular}{l | ccc}
\toprule
\multicolumn{4}{c}{LLM Code Generation and Refactoring Benchmark} \\
\midrule
\midrule
\multirow{2}{*}{\textbf{Model}} & Aider & \multirow{2}{*}{McEval} & HumanEval \\
& CodeEditing & & (0-shot) \\
\midrule
Qwen2.5-Coder-7B-I & 57.9 & 60.3 & 88.4 \\
Llama-3-8B-Instruct & 37.6 & 32.0 & 62.2 \\
DeepSeek-Coder-V2 LI & 48.9 & 54.7 & 81.1 \\
DeepSeek-Coder 33B I & 49.6 & 54.3 & 79.3 \\
CodeStral-22B & 48.1 & 50.5 & 78.1 \\
GPT-4o (240513) & 54.0 & 72.9 & 90.2 \\
\bottomrule
\end{tabular}
\caption{Evaluation of state-of-the-art large language models, with all values reported in percentages. McEval results represent Pass@1 
performance, while HumanEval scores reflect 0-shot capabilities. DeepSeek-Coder-V2 LI denotes the Lite Instruct variant, DeepSeek-Coder 33B I refers to the Instruct version, and Qwen2.5-Coder-7B-I indicates the Instruct variant.}
\label{tab:LLM_benchmarking}
\end{table}

Given the sensitivity of the data and the stringent data protection requirements in the automotive domain, our use case necessitates a locally deployable LLM capable of handling complex programming tasks without compromising data security. Among models with a token size of less than 10B, Qwen2.5-Coder-7B-Instruct proves as the most effective option, offering excellent performance while being the smallest in this category.

Once the LLM has generated the output, the next step is to extract the code and install the necessary libraries. This phase includes dependency management, ensuring all necessary libraries and tools are properly defined and integrated into the software environment. By automating these tasks, the framework accelerates the development cycle while maintaining precision and reliability. This structured approach to LLM-based generation bridges the gap between user-provided input and actionable software artefacts.

\subsection{Static Validation Module}
Once the code has been extracted from the LLM analysis, the first step in the evaluation process is static code analysis. This phase involves a series of checks to ensure the safety, functionality and adherence to design specifications of the generated code. Static code analysis plays a critical role in identifying potential problems early on, providing a strong foundation for seamless integration into the main framework.
To advance to the next state (\(S\)) and proceed to integration testing, the generated software (\(SW\)) must successfully pass a series of static checks (\(c_i\)). The process shown in \ref{algorithm1} involves an iterative cycle of software generation and refinement, where each iteration systematically resolves errors (\(E\)) identified during the previous analysis. Through this approach, the software progressively achieves compliance with the required safety and functional standards, ensuring readiness for the integration phase.

\begin{algorithm}[ht]
    \caption{Static Analysis and Error Handling}
    \label{algorithm1}
    \begin{algorithmic}[1]
        \While{$\exists \, s^s_{i} \in S \text{ such that } s^s_{i} \neq \text{success}$}
            \State $SW \gets \texttt{generate\_code}(s^s_{i}, E)$
            \For{$c_i \in C$}
                \State $s^s_{i} \gets \texttt{analyze}(SW, c_i)$
                \If{$s^s_{i} == \text{success}$}
                    \State \textbf{continue}
                \Else
                    \State $E \gets \texttt{get\_error\_analysis}(SW, c_i)$
                    \State \textbf{break}
                \EndIf
            \EndFor
        \EndWhile
        \State $S \gets \texttt{run\_integration\_monitoring}(SW)$
    \end{algorithmic}
\end{algorithm}

The static analysis process consists of several key checks:
It starts with a structure check, which verifies that the function name of the primary functionality is in the appropriate place in the generated code. This ensures compatibility with the broader framework and establishes a baseline for integration.

The compilation check in C++ ensures that the code can be successfully compiled by validating it. The compiler performs multiple checks, including syntactic verification to enforce correct grammar and semantic analysis to ensure meaningful execution. Modern compilers employ sophisticated multi-layered validation mechanisms to detect type inconsistencies, scope violations and improper memory usage. This process guarantees both syntactic and semantic correctness, reinforcing code reliability and robustness in high-performance computing environments.

Next, the static code style and design check enforces adherence to established formatting and design principles, ensuring maintainability and consistency. Tools such as cppcheck for C++ validate compliance with standards such as MISRA to address stylistic and semantic issues.

Finally, unit testing validates the functional behaviour of the code. Frameworks such as Google Test in C++ are used to run comprehensive test suites, covering diverse scenarios to confirm that inputs and outputs align with specified requirements. These tests serve as a safety net against regressions and ensure correctness in future iterations. When using unit testing as feedback, it is crucial to withhold specific failure details from the LLM, preventing over-fitting to individual tests and preserving generalizable behaviour.

Taken together, these steps constitute a rigorous and systematic approach to static analysis, ensuring that the code meets the highest standards of safety, compatibility and reliability, while laying a robust foundation for subsequent development and integration.

\subsection{Integration Monitoring Module}

\begin{figure}[h]
    \vspace{1em}
    \centering
    \includegraphics[width=0.48\textwidth]{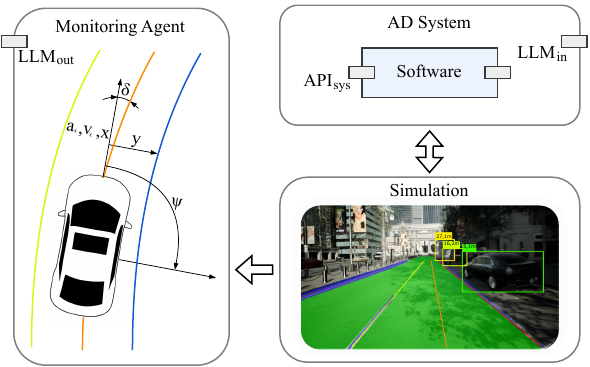}
    \caption{The integration monitoring framework consists of three key components: the monitoring agent, which acts as a client to the simulation server and has access to ground truth data; the simulation server, which provides synthetic sensor and vehicle state information; and the AD system, which provides an autonomous driving system that allows to integrate the generated software through clear APIs.}
    \label{fig:integration_monitoring}
\end{figure}

After successfully passing the static analysis phase and demonstrating full compliance with static safety requirements, the generated software is integrated into the Autonomous Driving (AD) system. This integration is achieved through the use of well-defined and strictly verified Application Programming Interfaces (APIs), which have been validated during the static analysis phase for consistency with the previously verified software components. These validated System APIs ($API_{sys}$) ensure a seamless and robust connection between the generated software and the broader AD system architecture. The AD system, which receives both vehicle state information and sensor data from a simulation environment, consists of the three core components of autonomous driving systems: detection, planning and control. For the detection module, YOLOP \cite{Wu2022} is used, enabling simultaneous semantic segmentation and object detection, providing essential contextual information. The CARLA simulator serves as the simulation environment, generating synthetic sensor data, such as camera feeds, as shown in Figure~\ref{fig:integration_monitoring}. To enrich the object detection with depth information, a depth camera is integrated to provide distance measurements relative to the ego vehicle. 

This data, along with other vehicle states obtained from the inertial measurement unit (IMU), is transmitted to the AD system. The integration monitoring framework includes a monitoring agent (see figure \ref{fig:integration_monitoring}) that uses a simple API to inspect the environment and system behaviour through its access to the ground truth data from the simulation server. User-provided system behaviour specifications serve as a fundamental element in this framework. These specifications articulate simple mathematical functions designed to validate vehicle behaviour against pre-defined performance criteria. The integration process follows a structured timeline that includes three distinct phases: initialization, data processing and evaluation. During the initialization phase, multi-processing is used to simultaneously start the simulation environment, the monitoring agent, and the automated driving (AD) system, ensuring that the performance of the generated function remains unaffected. Once operational, the integration monitoring system performs three critical functions: \texttt{get\_data\_from\_carla\_server}, \texttt{process\_carla\_data()} and \texttt{add\_data\_to\_statistics()}. At the end of a predefined integration test duration, the \texttt{evaluate\_data()} function is triggered to perform the mathematical validations defined in the system behaviour specifications. These validations span a spectrum of criteria, ranging from behavioural and comfort-related metrics to strict timing requirements. Upon successful completion of this stage, the generated software, refined through iterative interactions with the LLM, achieves the robustness and reliability required for seamless integration, meeting all safety-critical and performance criteria.


\section{Evaluation}
To evaluate the proposed framework, we generate a test function designed to demonstrate its effectiveness in a realistic scenario. Specifically, we select the adaptive cruise control (ACC) system, a widely studied application in the automotive domain. The ACC system offers robust evaluation capabilities in simulation and aligns with the stringent ISO standards, providing a benchmark for the performance and reliability of our framework.

\subsection{Simulation Environment and Hardware Setup}  
The primary objective of the ACC system is to control the longitudinal motion of a vehicle. As a critical component of the control subsystem in autonomous driving systems, the ACC takes as input the bounding box of the lead vehicle including distance information and the inertial measurement unit (IMU) data of the ego vehicle. Using these inputs, the system calculates the necessary longitudinal motion and generates throttle and brake commands as outputs. These interfaces are essential for maintaining the correct vehicle speed and spacing in dynamic driving conditions. To ensure seamless integration within the integration system, the AD system includes additional modules required for vehicle operation. These include object detection and segmentation to identify surrounding objects, as well as motion planning and a lateral control module to ensure coordinated vehicle maneuvers. The simulation environment replicates real-world driving scenarios, providing a controlled yet dynamic setting to evaluate the ACC function. The system is tested under various conditions to validate its robustness, efficiency and compliance with ISO requirements. The hardware setup for this evaluation shown in Table \ref{tab:hardware} is critical to achieve high-performance execution and accurate real-time simulation.

\begin{table}[h]
\centering
\begin{tabular}{l|l}
\toprule
\multicolumn{2}{c}{Hardware Setup for the Generation and Testing Modules} \\
\midrule
\midrule
\textbf{CPU} & AMD Ryzen 9 9950X (16x 4.3 GHz, 170W) \\
\textbf{GPU} & 2 x NVIDIA GeForce RTX 4090 24GB \\
\textbf{RAM} & 64GB DDR5-5600 Vengeance (2x 32GB) \\
\bottomrule
\end{tabular}
\caption{To accelerate the development process, we leverage a high-performance hardware configuration to run the LLM locally and execute the algorithm within the simulation environment faster than real-time, ensuring efficient experimentation and iterative refinement.}
\label{tab:hardware}
\end{table}

To enhance computational efficiency, we use accelerated inference techniques to reduce latency and enable faster processing. Memory management optimizations are also implemented to minimize gaps in GPU memory allocation. By dynamically growing memory segments, the framework ensures efficient utilization of VRAM, reducing fragmentation and improving stability during prolonged simulations. 

\subsection{Requirements Engineering and System Specification}

The requirements for the evaluation of the Adaptive Cruise Control (ACC) function are derived from a combination of multiple sources. The primary standard used is ISO 15622\cite{ISO15622}, which specifies the performance, safety and functional behaviour requirements for ACC systems. This standard serves as the foundation for defining the control strategy and minimum functional requirements of ACC systems, including parameters such as time gap ($\tau$), clearance ($c$) and ego vehicle speed ($v$).

\begin{figure}[h]
\vspace{1em}
\centering
\includegraphics[width=0.48\textwidth]{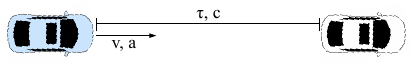}
\caption{ISO 15622 (Intelligent Transport Systems: Adaptive cruise control systems \textendash{} Performance requirements and test procedures) defines the basic control strategy and minimum functional requirements for ACC systems. The time gap is introduced as $\tau$, the clearance as $c$ and the ego vehicle speed as $v$.}
\label{fig:acc_system}
\end{figure}

From this, we derive the requirement that the minimum clearance should satisfy:
\begin{equation}
\text{MAX}(c_{\text{min}}, \tau_{\text{min}} \cdot v)
\end{equation}
In addition, we define a nominal following distance of 10 meters as an optimal balance between safety and driving comfort. In addition, to ensure physically possible driving manoeuvres, we set the requirement to reduce the maximum possible acceleration (positive and negative). The maximum acceleration is therefore defined as:

\begin{equation}
|a| < 5 m/s^{2}
\end{equation}

if no emergency brake is applied. The architecture, interfaces and operating logic are carefully defined, following established best practice in control system development. Generative AI is given controlled access to throttle and brake APIs, complemented by real-time vehicle state data and object detection results, enabling seamless integration with the wider autonomous driving (AD) system. Software quality and reliability is ensured by MISRA-compliant static analysis, which facilitates early detection of potential problems. In addition, preconditions for integration monitoring, as outlined in ISO 15622, form the foundation for subsequent test phases, ensuring thorough evaluation and compliance with critical standards.

\subsection{Evaluating the Performance of the generated ACC System}
\begin{figure}[t]
    \vspace{1em}
    \centering
    \includegraphics[width=0.48\textwidth]{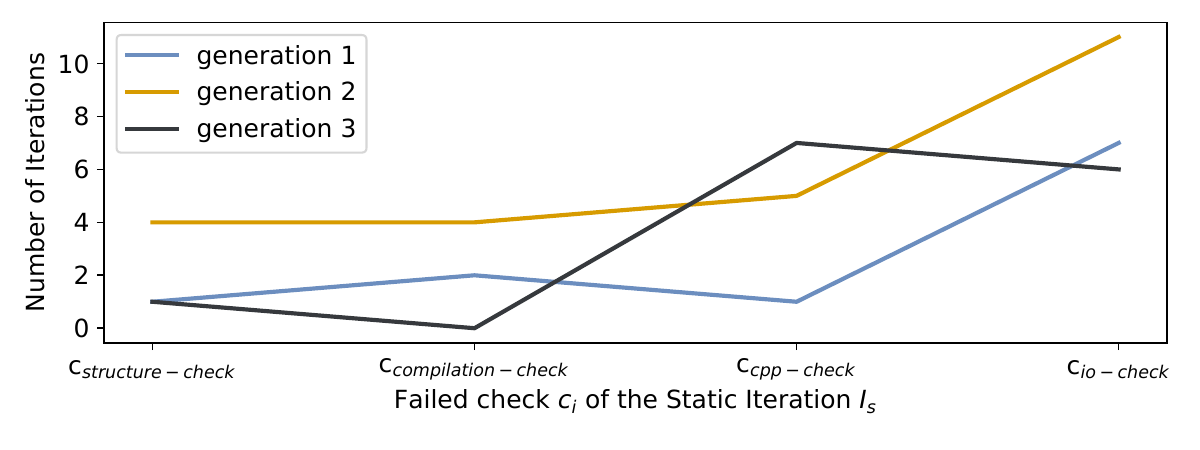}
    \caption{Across three independent code generation runs using Qwen2.5-Coder-7B-I, an average of 16.3 iterations were required to pass all static tests. Structure and compilation checks failed the least (2 iterations on average). While the CPP check, including MISRA compliance, took an average of 4.3 iterations, the most demanding phase was unit testing, which required 8 iterations on average to ensure correctness.}
    \label{fig:static_evaluation}
\end{figure}

To analyze the behavior of the LLM across both software states (Figure \ref{fig:safe_states}), we evaluate its ability to generate code that meets the requirements of the verified state (static checks) and the safe state (combined static and integration checks).
We apply the defined requirement checks and initiate code generation, running three independent iterations to analyse and compare the number of attempts it takes the LLM framework to produce code that satisfies the static requirements of a given check (Figure \ref{fig:static_evaluation}).

\begin{figure}[t]
    \vspace{1em}
    \centering
    \includegraphics[width=0.48\textwidth]{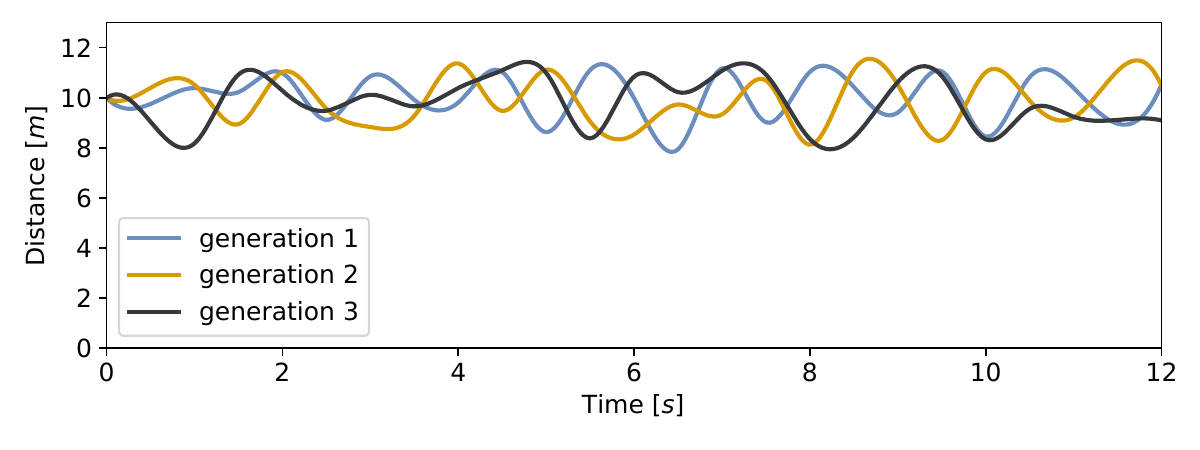}
    \caption{The distance between the ego vehicle and the leading vehicle over time shows that all three generated functions maintain a consistent following distance of approximately 10 metres. The behaviour shows only minor differences across all implementations, with adjustments occurring mainly in the 8m to 12m range.}
    \label{fig:distance}
\end{figure}

\begin{figure}[t]
    \vspace{1em}
    \centering
    \includegraphics[width=0.48\textwidth]{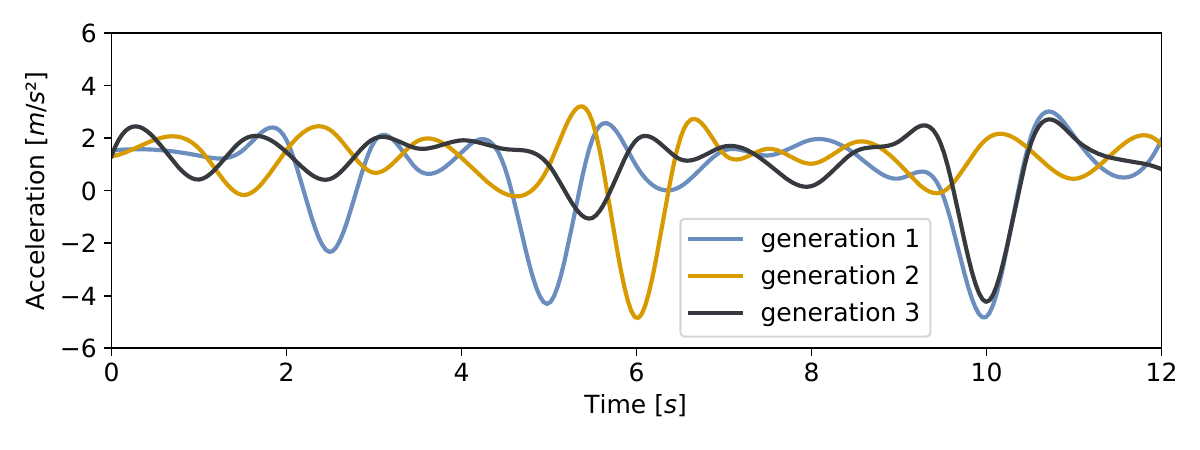}
    \caption{The acceleration profiles over time show the typical oscillatory behaviour of all the functions evaluated. Notably, each function adheres to the predefined safety requirement, ensuring that the braking does not exceed the threshold of 5 m/s².}
    \label{fig:acceleration}
\end{figure}

To ensure effective testing, the behaviour of the leading vehicle is randomised during integration monitoring. This approach prevents the LLM from tailoring the generated function to a specific scenario, thereby improving generalization. The results, shown in figures \ref{fig:distance} and \ref{fig:acceleration}, demonstrate consistent compliance with the defined requirements. The generated functions maintain the expected vehicle behaviour, and even under emergency conditions the deceleration remains within the prescribed safety threshold.


\section{CONCLUSION AND FUTURE WORK}

This work addresses the problem of integrating generative AI into safety-critical automotive software development. Specifically, this work aims to (1) develop an LLM-driven software development framework that ensures compliance with functional safety requirements and (2) establish a structured pipeline for validation through static code analysis and integration monitoring. By using structured specifications, automated refinement, and iterative validation, the proposed framework enables efficient and reliable software generation for safety-critical applications. Using a case study on adaptive cruise control (ACC) case study, our experiments demonstrate that the generated functions meet predefined safety and performance constraints, maintaining safe following distances and adhering to acceleration limits even in emergency scenarios.

Future work will extend the framework by adapting mathematical guarantees to prove software correctness. Additional extensions could involve increasing the number and complexity of requirements to better assess the robustness of various LLMs, as well as evaluating their performance across different automotive functions. Furthermore, the exploration of system-level AI-driven software design methodologies could facilitate the development of a structured model that improves the integration of generative AI into the automotive software development lifecycle, building upon and refining the ASPICE standard. Further evaluation in real-world testing will also be critical to validate the effectiveness of the framework in practical deployment scenarios.


\printbibliography

@book{Dijkstra1997,
author = {Dijkstra, Edsger Wybe},
title = {A  Discipline of Programming},
year = {1997},
isbn = {013215871X},
publisher = {Prentice Hall PTR},
address = {USA},
edition = {1st}
}

@book{sommerville2011,
author = {Sommerville, Ian},
title = {Software Engineering},
year = {2011},
edition = {9th},
publisher = {Addison-Wesley},
address = {Boston, MA, USA}
}

@book{pressman2014,
author = {Pressman, Roger S.},
title = {Software Engineering: A Practitioner's Approach},
year = {2014},
edition = {8th},
publisher = {McGraw-Hill},
address = {New York, NY, USA}
}

@book{beck2003,
author = {Beck, Kent},
title = {Test-Driven Development by Example},
year = {2003},
publisher = {Addison-Wesley},
address = {Boston, MA, USA}
}

@book{silberschatz2020,
author = {Silberschatz, Abraham and Korth, Henry F. and Sudarshan, S.},
title = {Database System Concepts},
year = {2020},
edition = {7th},
publisher = {McGraw-Hill},
address = {New York, NY, USA}
}

@book{bass2012,
author = {Bass, Len and Clements, Paul and Kazman, Rick},
title = {Software Architecture in Practice},
year = {2012},
edition = {3rd},
publisher = {Addison-Wesley},
address = {Boston, MA, USA}
}

@book{higham2002,
author = {Higham, Nicholas J.},
title = {Accuracy and Stability of Numerical Algorithms},
year = {2002},
edition = {2nd},
publisher = {Society for Industrial and Applied Mathematics (SIAM)},
address = {Philadelphia, PA, USA}
}

@book{fowler2004,
author = {Fowler, Martin},
title = {Patterns of Enterprise Application Architecture},
year = {2004},
publisher = {Addison-Wesley},
address = {Boston, MA, USA}
}

@book{shavit2012,
author = {Shavit, Nir and Herlihy, Maurice},
title = {The Art of Multiprocessor Programming},
year = {2012},
publisher = {Morgan Kaufmann},
address = {San Francisco, CA, USA}
}

@book{viega2001,
author = {Viega, John and McGraw, Gary},
title = {Building Secure Software: How to Avoid Security Problems the Right Way},
year = {2001},
publisher = {Addison-Wesley},
address = {Boston, MA, USA}
}

@book{ISO26262,
  title = {ISO 26262: Road vehicles – Functional safety},
  author = {{International Organization for Standardization}},
  year = {2018},
  publisher = {ISO},
  address = {Geneva, Switzerland}
}

@book{ASPICE2022,
  title = {Automotive SPICE Process Assessment Model},
  author = {{VDA QMC Working Group 13}},
  year = {2022},
  publisher = {German Association of the Automotive Industry (VDA)},
  address = {Berlin, Germany}
}

@book{MISRA2012,
  title = {MISRA C: Guidelines for the use of the C language in critical systems},
  author = {{Motor Industry Software Reliability Association}},
  year = {2012},
  edition = {3rd},
  publisher = {MISRA},
  address = {UK}
}

@online{GoogleTest,
  title = {Google Test: C++ Testing Framework},
  author = {{Google}},
  year = {2023},
  url = {https://github.com/google/googletest},
  note = {Accessed: 2023-12-10}
}

@misc{vaswani2023,
      title={Attention Is All You Need}, 
      author={Ashish Vaswani and Noam Shazeer and Niki Parmar and Jakob Uszkoreit and Llion Jones and Aidan N. Gomez and Lukasz Kaiser and Illia Polosukhin},
      year={2023},
      eprint={1706.03762},
      archivePrefix={arXiv},
      primaryClass={cs.CL},
      url={https://arxiv.org/abs/1706.03762}, 
}

@article{hui2024,
      title={Qwen2. 5-Coder Technical Report},
      author={Hui, Binyuan and Yang, Jian and Cui, Zeyu and Yang, Jiaxi and Liu, Dayiheng and Zhang, Lei and Liu, Tianyu and Zhang, Jiajun and Yu, Bowen and Dang, Kai and others},
      journal={arXiv preprint arXiv:2409.12186},
      year={2024}
}

@inbook{Scius2024,
   title={Zero-Shot Prompting and Few-Shot Fine-Tuning: Revisiting Document Image Classification Using Large Language Models},
   ISBN={9783031784958},
   ISSN={1611-3349},
   DOI={10.1007/978-3-031-78495-8_10},
   booktitle={Pattern Recognition},
   publisher={Springer Nature Switzerland},
   author={Scius-Bertrand, Anna and Jungo, Michael and Vögtlin, Lars and Spat, Jean-Marc and Fischer, Andreas},
   year={2024},
   month=dec, pages={152–166} }

@misc{zhang2024,
      title={Multimodal Chain-of-Thought Reasoning in Language Models}, 
      author={Zhuosheng Zhang and Aston Zhang and Mu Li and Hai Zhao and George Karypis and Alex Smola},
      year={2024},
      eprint={2302.00923},
      archivePrefix={arXiv},
      primaryClass={cs.CL},
}

@misc{kong2024,
      title={Better Zero-Shot Reasoning with Role-Play Prompting}, 
      author={Aobo Kong and Shiwan Zhao and Hao Chen and Qicheng Li and Yong Qin and Ruiqi Sun and Xin Zhou and Enzhi Wang and Xiaohang Dong},
      year={2024},
      eprint={2308.07702},
      archivePrefix={arXiv},
      primaryClass={cs.CL},
      url={https://arxiv.org/abs/2308.07702}, 
}

@misc{grattafiori2024,
      title={The Llama 3 Herd of Models}, 
      author={Aaron Grattafiori and Abhimanyu Dubey and Abhinav Jauhri and Abhinav Pandey and Abhishek Kadian and Ahmad Al-Dahle and Aiesha Letman and Akhil Mathur and Alan Schelten and Alex Vaughan and Amy Yang and Angela Fan and Anirudh Goyal and Anthony Hartshorn and Aobo Yang and Archi Mitra and Archie Sravankumar and Artem Korenev and Arthur Hinsvark and Arun Rao and Aston Zhang and Aurelien Rodriguez and Austen Gregerson and Ava Spataru and Baptiste Roziere and Bethany Biron and Binh Tang and Bobbie Chern and Charlotte Caucheteux and Chaya Nayak and Chloe Bi and Chris Marra and Chris McConnell and Christian Keller and Christophe Touret and Chunyang Wu and Corinne Wong and Cristian Canton Ferrer and Cyrus Nikolaidis and Damien Allonsius and Daniel Song and Danielle Pintz and Danny Livshits and Danny Wyatt and David Esiobu and Dhruv Choudhary and Dhruv Mahajan and Diego Garcia-Olano and Diego Perino and Dieuwke Hupkes and Egor Lakomkin and Ehab AlBadawy and Elina Lobanova and Emily Dinan and Eric Michael Smith and Filip Radenovic and Francisco Guzmán and Frank Zhang and Gabriel Synnaeve and Gabrielle Lee and Georgia Lewis Anderson and Govind Thattai and Graeme Nail and Gregoire Mialon and Guan Pang and Guillem Cucurell and Hailey Nguyen and Hannah Korevaar and Hu Xu and Hugo Touvron and Iliyan Zarov and Imanol Arrieta Ibarra and Isabel Kloumann and Ishan Misra and Ivan Evtimov and Jack Zhang and Jade Copet and Jaewon Lee and Jan Geffert and Jana Vranes and Jason Park and Jay Mahadeokar and Jeet Shah and Jelmer van der Linde and Jennifer Billock and Jenny Hong and Jenya Lee and Jeremy Fu and Jianfeng Chi and Jianyu Huang and Jiawen Liu and Jie Wang and Jiecao Yu and Joanna Bitton and Joe Spisak and Jongsoo Park and Joseph Rocca and Joshua Johnstun and Joshua Saxe and Junteng Jia and Kalyan Vasuden Alwala and Karthik Prasad and Kartikeya Upasani and Kate Plawiak and Ke Li and Kenneth Heafield and Kevin Stone and Khalid El-Arini and Krithika Iyer and Kshitiz Malik and Kuenley Chiu and Kunal Bhalla and Kushal Lakhotia and Lauren Rantala-Yeary and Laurens van der Maaten and Lawrence Chen and Liang Tan and Liz Jenkins and Louis Martin and Lovish Madaan and Lubo Malo and Lukas Blecher and Lukas Landzaat and Luke de Oliveira and Madeline Muzzi and Mahesh Pasupuleti and Mannat Singh and Manohar Paluri and Marcin Kardas and Maria Tsimpoukelli and Mathew Oldham and Mathieu Rita and Maya Pavlova and Melanie Kambadur and Mike Lewis and Min Si and Mitesh Kumar Singh and Mona Hassan and Naman Goyal and Narjes Torabi and Nikolay Bashlykov and Nikolay Bogoychev and Niladri Chatterji and Ning Zhang and Olivier Duchenne and Onur Çelebi and Patrick Alrassy and Pengchuan Zhang and Pengwei Li and Petar Vasic and Peter Weng and Prajjwal Bhargava and Pratik Dubal and Praveen Krishnan and Punit Singh Koura and Puxin Xu and Qing He and Qingxiao Dong and Ragavan Srinivasan and Raj Ganapathy and Ramon Calderer and Ricardo Silveira Cabral and Robert Stojnic and Roberta Raileanu and Rohan Maheswari and Rohit Girdhar and Rohit Patel and Romain Sauvestre and Ronnie Polidoro and Roshan Sumbaly and Ross Taylor and Ruan Silva and Rui Hou and Rui Wang and Saghar Hosseini and Sahana Chennabasappa and Sanjay Singh and Sean Bell and Seohyun Sonia Kim and Sergey Edunov and Shaoliang Nie and Sharan Narang and Sharath Raparthy and Sheng Shen and Shengye Wan and Shruti Bhosale and Shun Zhang and Simon Vandenhende and Soumya Batra and Spencer Whitman and Sten Sootla and Stephane Collot and Suchin Gururangan and Sydney Borodinsky and Tamar Herman and Tara Fowler and Tarek Sheasha and Thomas Georgiou and Thomas Scialom and Tobias Speckbacher and Todor Mihaylov and Tong Xiao and Ujjwal Karn and Vedanuj Goswami and Vibhor Gupta and Vignesh Ramanathan and Viktor Kerkez and Vincent Gonguet and Virginie Do and Vish Vogeti and Vítor Albiero and Vladan Petrovic and Weiwei Chu and Wenhan Xiong and Wenyin Fu and Whitney Meers and Xavier Martinet and Xiaodong Wang and Xiaofang Wang and Xiaoqing Ellen Tan and Xide Xia and Xinfeng Xie and Xuchao Jia and Xuewei Wang and Yaelle Goldschlag and Yashesh Gaur and Yasmine Babaei and Yi Wen and Yiwen Song and Yuchen Zhang and Yue Li and Yuning Mao and Zacharie Delpierre Coudert and Zheng Yan and Zhengxing Chen and Zoe Papakipos and Aaditya Singh and Aayushi Srivastava and Abha Jain and Adam Kelsey and Adam Shajnfeld and Adithya Gangidi and Adolfo Victoria and Ahuva Goldstand and Ajay Menon and Ajay Sharma and Alex Boesenberg and Alexei Baevski and Allie Feinstein and Amanda Kallet and Amit Sangani and Amos Teo and Anam Yunus and Andrei Lupu and Andres Alvarado and Andrew Caples and Andrew Gu and Andrew Ho and Andrew Poulton and Andrew Ryan and Ankit Ramchandani and Annie Dong and Annie Franco and Anuj Goyal and Aparajita Saraf and Arkabandhu Chowdhury and Ashley Gabriel and Ashwin Bharambe and Assaf Eisenman and Azadeh Yazdan and Beau James and Ben Maurer and Benjamin Leonhardi and Bernie Huang and Beth Loyd and Beto De Paola and Bhargavi Paranjape and Bing Liu and Bo Wu and Boyu Ni and Braden Hancock and Bram Wasti and Brandon Spence and Brani Stojkovic and Brian Gamido and Britt Montalvo and Carl Parker and Carly Burton and Catalina Mejia and Ce Liu and Changhan Wang and Changkyu Kim and Chao Zhou and Chester Hu and Ching-Hsiang Chu and Chris Cai and Chris Tindal and Christoph Feichtenhofer and Cynthia Gao and Damon Civin and Dana Beaty and Daniel Kreymer and Daniel Li and David Adkins and David Xu and Davide Testuggine and Delia David and Devi Parikh and Diana Liskovich and Didem Foss and Dingkang Wang and Duc Le and Dustin Holland and Edward Dowling and Eissa Jamil and Elaine Montgomery and Eleonora Presani and Emily Hahn and Emily Wood and Eric-Tuan Le and Erik Brinkman and Esteban Arcaute and Evan Dunbar and Evan Smothers and Fei Sun and Felix Kreuk and Feng Tian and Filippos Kokkinos and Firat Ozgenel and Francesco Caggioni and Frank Kanayet and Frank Seide and Gabriela Medina Florez and Gabriella Schwarz and Gada Badeer and Georgia Swee and Gil Halpern and Grant Herman and Grigory Sizov and Guangyi and Zhang and Guna Lakshminarayanan and Hakan Inan and Hamid Shojanazeri and Han Zou and Hannah Wang and Hanwen Zha and Haroun Habeeb and Harrison Rudolph and Helen Suk and Henry Aspegren and Hunter Goldman and Hongyuan Zhan and Ibrahim Damlaj and Igor Molybog and Igor Tufanov and Ilias Leontiadis and Irina-Elena Veliche and Itai Gat and Jake Weissman and James Geboski and James Kohli and Janice Lam and Japhet Asher and Jean-Baptiste Gaya and Jeff Marcus and Jeff Tang and Jennifer Chan and Jenny Zhen and Jeremy Reizenstein and Jeremy Teboul and Jessica Zhong and Jian Jin and Jingyi Yang and Joe Cummings and Jon Carvill and Jon Shepard and Jonathan McPhie and Jonathan Torres and Josh Ginsburg and Junjie Wang and Kai Wu and Kam Hou U and Karan Saxena and Kartikay Khandelwal and Katayoun Zand and Kathy Matosich and Kaushik Veeraraghavan and Kelly Michelena and Keqian Li and Kiran Jagadeesh and Kun Huang and Kunal Chawla and Kyle Huang and Lailin Chen and Lakshya Garg and Lavender A and Leandro Silva and Lee Bell and Lei Zhang and Liangpeng Guo and Licheng Yu and Liron Moshkovich and Luca Wehrstedt and Madian Khabsa and Manav Avalani and Manish Bhatt and Martynas Mankus and Matan Hasson and Matthew Lennie and Matthias Reso and Maxim Groshev and Maxim Naumov and Maya Lathi and Meghan Keneally and Miao Liu and Michael L. Seltzer and Michal Valko and Michelle Restrepo and Mihir Patel and Mik Vyatskov and Mikayel Samvelyan and Mike Clark and Mike Macey and Mike Wang and Miquel Jubert Hermoso and Mo Metanat and Mohammad Rastegari and Munish Bansal and Nandhini Santhanam and Natascha Parks and Natasha White and Navyata Bawa and Nayan Singhal and Nick Egebo and Nicolas Usunier and Nikhil Mehta and Nikolay Pavlovich Laptev and Ning Dong and Norman Cheng and Oleg Chernoguz and Olivia Hart and Omkar Salpekar and Ozlem Kalinli and Parkin Kent and Parth Parekh and Paul Saab and Pavan Balaji and Pedro Rittner and Philip Bontrager and Pierre Roux and Piotr Dollar and Polina Zvyagina and Prashant Ratanchandani and Pritish Yuvraj and Qian Liang and Rachad Alao and Rachel Rodriguez and Rafi Ayub and Raghotham Murthy and Raghu Nayani and Rahul Mitra and Rangaprabhu Parthasarathy and Raymond Li and Rebekkah Hogan and Robin Battey and Rocky Wang and Russ Howes and Ruty Rinott and Sachin Mehta and Sachin Siby and Sai Jayesh Bondu and Samyak Datta and Sara Chugh and Sara Hunt and Sargun Dhillon and Sasha Sidorov and Satadru Pan and Saurabh Mahajan and Saurabh Verma and Seiji Yamamoto and Sharadh Ramaswamy and Shaun Lindsay and Shaun Lindsay and Sheng Feng and Shenghao Lin and Shengxin Cindy Zha and Shishir Patil and Shiva Shankar and Shuqiang Zhang and Shuqiang Zhang and Sinong Wang and Sneha Agarwal and Soji Sajuyigbe and Soumith Chintala and Stephanie Max and Stephen Chen and Steve Kehoe and Steve Satterfield and Sudarshan Govindaprasad and Sumit Gupta and Summer Deng and Sungmin Cho and Sunny Virk and Suraj Subramanian and Sy Choudhury and Sydney Goldman and Tal Remez and Tamar Glaser and Tamara Best and Thilo Koehler and Thomas Robinson and Tianhe Li and Tianjun Zhang and Tim Matthews and Timothy Chou and Tzook Shaked and Varun Vontimitta and Victoria Ajayi and Victoria Montanez and Vijai Mohan and Vinay Satish Kumar and Vishal Mangla and Vlad Ionescu and Vlad Poenaru and Vlad Tiberiu Mihailescu and Vladimir Ivanov and Wei Li and Wenchen Wang and Wenwen Jiang and Wes Bouaziz and Will Constable and Xiaocheng Tang and Xiaojian Wu and Xiaolan Wang and Xilun Wu and Xinbo Gao and Yaniv Kleinman and Yanjun Chen and Ye Hu and Ye Jia and Ye Qi and Yenda Li and Yilin Zhang and Ying Zhang and Yossi Adi and Youngjin Nam and Yu and Wang and Yu Zhao and Yuchen Hao and Yundi Qian and Yunlu Li and Yuzi He and Zach Rait and Zachary DeVito and Zef Rosnbrick and Zhaoduo Wen and Zhenyu Yang and Zhiwei Zhao and Zhiyu Ma},
      year={2024},
      eprint={2407.21783},
      archivePrefix={arXiv},
      primaryClass={cs.AI},
      url={https://arxiv.org/abs/2407.21783}, 
}

@misc{dosovitskiy2017,
      title={CARLA: An Open Urban Driving Simulator}, 
      author={Alexey Dosovitskiy and German Ros and Felipe Codevilla and Antonio Lopez and Vladlen Koltun},
      year={2017},
      eprint={1711.03938},
      archivePrefix={arXiv},
      primaryClass={cs.LG},
      url={https://arxiv.org/abs/1711.03938}, 
}

@phdthesis{Eriksson2011, series={Kandidatexjobb CSC}, title={Comparison between JSON and YAML for Data Serialization.}, url={https://urn.kb.se/resolve?urn=urn:nbn:se:kth:diva-130815}, abstractNote={This report determines and discusses the primary differences between two different serialization formats, namely YAML and JSON. A general introduction to the concepts of serialization and parsing is provided first, which also explains how they can be used to transfer and store data. This is followed by an analysis of the YAML and JSON formats, where functionality, primary use cases, and syntax is described. In addition to this the percieved performance of implementations for both formats will also be investigated by conducting a number of tests. Using the combined background information and results from the tests, conclusions regarding the main differences between the two are then determined and discussed. }, author={Eriksson, Malin and Hallberg, Victor}, year={2011}, collection={Kandidatexjobb CSC} }

@article{Wu2022,
   title={YOLOP: You Only Look Once for Panoptic Driving Perception},
   volume={19},
   ISSN={2731-5398},
   url={http://dx.doi.org/10.1007/s11633-022-1339-y},
   DOI={10.1007/s11633-022-1339-y},
   number={6},
   journal={Machine Intelligence Research},
   publisher={Springer Science and Business Media LLC},
   author={Wu, Dong and Liao, Man-Wen and Zhang, Wei-Tian and Wang, Xing-Gang and Bai, Xiang and Cheng, Wen-Qing and Liu, Wen-Yu},
   year={2022},
   month=nov, pages={550–562} }

@misc{ISO15622,
  title        = {Intelligent transport systems — Adaptive Cruise Control systems — Performance requirements and test procedures},
  author       = {{International Organization for Standardization}},
  year         = {2018},
  number       = {ISO 15622:2018},
  url          = {https://www.iso.org/standard/71515.html},
  publisher    = {International Organization for Standardization}
}

@misc{mceval2024,
      title={McEval: Massively Multilingual Code Evaluation}, 
      author={Linzheng Chai and Shukai Liu and Jian Yang and Yuwei Yin and Ke Jin and Jiaheng Liu and Tao Sun and Ge Zhang and Changyu Ren and Hongcheng Guo and Zekun Wang and Boyang Wang and Xianjie Wu and Bing Wang and Tongliang Li and Liqun Yang and Sufeng Duan and Zhoujun Li},
      year={2024},
      eprint={2406.07436},
      archivePrefix={arXiv},
      primaryClass={cs.PL},
}

@misc{aider2024,
  title={{Aider Code Editing Benchmark}},
  author={{Aider-AI}},
  year={2024},
  howpublished={\url{https://github.com/Aider-AI/aider/blob/main/benchmark/README.md}},
  note={Accessed: 2025-01-22}
}

@misc{gpt2020,
      title={Language Models are Few-Shot Learners}, 
      author={Tom B. Brown and Benjamin Mann and Nick Ryder and Melanie Subbiah and Jared Kaplan and Prafulla Dhariwal and Arvind Neelakantan and Pranav Shyam and Girish Sastry and Amanda Askell and Sandhini Agarwal and Ariel Herbert-Voss and Gretchen Krueger and Tom Henighan and Rewon Child and Aditya Ramesh and Daniel M. Ziegler and Jeffrey Wu and Clemens Winter and Christopher Hesse and Mark Chen and Eric Sigler and Mateusz Litwin and Scott Gray and Benjamin Chess and Jack Clark and Christopher Berner and Sam McCandlish and Alec Radford and Ilya Sutskever and Dario Amodei},
      year={2020},
      eprint={2005.14165},
      archivePrefix={arXiv},
      primaryClass={cs.CL},
      url={https://arxiv.org/abs/2005.14165}, 
}

@book{pressman2009,
author = {Pressman, Roger},
title = {Software Engineering: A Practitioner's Approach},
year = {2009},
isbn = {0073375977},
publisher = {McGraw-Hill, Inc.},
address = {USA},
edition = {7}
}

@INPROCEEDINGS{serban2018,
  author={Serban, Alexandru Constantin and Poll, Erik and Visser, Joost},
  booktitle={2018 IEEE International Conference on Software Architecture Companion (ICSA-C)}, 
  title={A Standard Driven Software Architecture for Fully Autonomous Vehicles}, 
  year={2018},
  volume={},
  number={},
  pages={120-127},
  doi={10.1109/ICSA-C.2018.00040}}

@book{Stroustrup2013,
author = {Stroustrup, Bjarne},
title = {The C++ Programming Language},
year = {2013},
isbn = {0321563840},
publisher = {Addison-Wesley Professional},
edition = {4th}
}

@misc{chen2021,
      title={Evaluating Large Language Models Trained on Code}, 
      author={Mark Chen and Jerry Tworek and Heewoo Jun and Qiming Yuan and Henrique Ponde de Oliveira Pinto and Jared Kaplan and Harri Edwards and Yuri Burda and Nicholas Joseph and Greg Brockman and Alex Ray and Raul Puri and Gretchen Krueger and Michael Petrov and Heidy Khlaaf and Girish Sastry and Pamela Mishkin and Brooke Chan and Scott Gray and Nick Ryder and Mikhail Pavlov and Alethea Power and Lukasz Kaiser and Mohammad Bavarian and Clemens Winter and Philippe Tillet and Felipe Petroski Such and Dave Cummings and Matthias Plappert and Fotios Chantzis and Elizabeth Barnes and Ariel Herbert-Voss and William Hebgen Guss and Alex Nichol and Alex Paino and Nikolas Tezak and Jie Tang and Igor Babuschkin and Suchir Balaji and Shantanu Jain and William Saunders and Christopher Hesse and Andrew N. Carr and Jan Leike and Josh Achiam and Vedant Misra and Evan Morikawa and Alec Radford and Matthew Knight and Miles Brundage and Mira Murati and Katie Mayer and Peter Welinder and Bob McGrew and Dario Amodei and Sam McCandlish and Ilya Sutskever and Wojciech Zaremba},
      year={2021},
      eprint={2107.03374},
      archivePrefix={arXiv},
      primaryClass={cs.LG},
}

@misc{deepseekai2024,
      title={DeepSeek-Coder-V2: Breaking the Barrier of Closed-Source Models in Code Intelligence}, 
      author={DeepSeek-AI and Qihao Zhu and Daya Guo and Zhihong Shao and Dejian Yang and Peiyi Wang and Runxin Xu and Y. Wu and Yukun Li and Huazuo Gao and Shirong Ma and Wangding Zeng and Xiao Bi and Zihui Gu and Hanwei Xu and Damai Dai and Kai Dong and Liyue Zhang and Yishi Piao and Zhibin Gou and Zhenda Xie and Zhewen Hao and Bingxuan Wang and Junxiao Song and Deli Chen and Xin Xie and Kang Guan and Yuxiang You and Aixin Liu and Qiushi Du and Wenjun Gao and Xuan Lu and Qinyu Chen and Yaohui Wang and Chengqi Deng and Jiashi Li and Chenggang Zhao and Chong Ruan and Fuli Luo and Wenfeng Liang},
      year={2024},
      eprint={2406.11931},
      archivePrefix={arXiv},
      primaryClass={cs.SE},
}

@misc{guo2024,
      title={DeepSeek-Coder: When the Large Language Model Meets Programming -- The Rise of Code Intelligence}, 
      author={Daya Guo and Qihao Zhu and Dejian Yang and Zhenda Xie and Kai Dong and Wentao Zhang and Guanting Chen and Xiao Bi and Y. Wu and Y. K. Li and Fuli Luo and Yingfei Xiong and Wenfeng Liang},
      year={2024},
      eprint={2401.14196},
      archivePrefix={arXiv},
      primaryClass={cs.SE},
}

@misc{codestral2024,
  title={Codestral: Hello, World!},
  author={Mistral AI},
  year={2024},
  url={https://mistral.ai/news/codestral/},
  note={Accessed: January 27, 2025}
}

@misc{cppcheck2024,
  author = {Daniel Marjamäki},
  title = {Cppcheck: A static analysis tool for C++},
  year = {2024},
  url = {https://cppcheck.sourceforge.io/}
}

@misc{sevenhuijsen2025,
      title={VeCoGen: Automating Generation of Formally Verified C Code with Large Language Models}, 
      author={Merlijn Sevenhuijsen and Khashayar Etemadi and Mattias Nyberg},
      year={2025},
      eprint={2411.19275},
      archivePrefix={arXiv},
      primaryClass={cs.SE},
      url={https://arxiv.org/abs/2411.19275}, 
}

\end{document}